\documentclass[pra,aps,showpacs,twocolumn,floatfix,superscriptaddress]{revtex4}  
\usepackage{graphicx}
\usepackage[ansinew]{inputenc}
\usepackage{array}
\usepackage{color}
\usepackage{amsmath}
\usepackage{amsxtra}
\usepackage{amstext}
\usepackage{amssymb}
\usepackage{latexsym}
\usepackage{dsfont}
\usepackage{verbatim}
\usepackage{comment}
\usepackage{epstopdf}

\begin{document}

\title
{Mutual emergence of noncausal optical response and nonclassicality in an optomechanical system }
\author{Devrim Tarhan}
\affiliation{Department of Computer Education and
Instructional Technology, Dumlup{\i}nar
University, K\"{u}tahya, Turkey}
\author{Mehmet Emre Ta\c{s}g\i n}
\affiliation{Institute of Nuclear Sciences, Hacettepe University, 06800, Ankara, Turkey}
\affiliation{to whom correspondence should be addressed}
\affiliation{Email: metasgin@hacettepe.edu.tr}
\date{\today}

\begin{abstract}
We show that single-mode nonclassicality of the output of an optomechanical cavity and the noncausal linear optical response of this cavity emerge at the same critical cavity-mechanical coupling. In other words, single-mode nonclassicality emerges when the barrier (in electromagnetism) avoiding faster-than-light communication is lifted off. The nature of the emergence of noncausal behavior does not depend on the length (boundary conditions) and the type of the cavity. Origin of the noncausal behavior is the temporal/frequency relations between the incident and reflected waves at the outer surface of the cavity. We further discuss the relations with the recent studies; (i) equivalence of the entanglement among identical particles to the nonclassicality of their quasiparticle excitations, (ii) necessity of superfluid behavior of vacuum, and (iii) entanglement-wormhole equivalence. 
\end{abstract}



\maketitle

\section{Introduction}

Aharonov {\it et al.} classified the emergence of superluminosity into two distinct cases \cite{SusskindSuperluminal}, more than four decades ago. They showed that field (theory) equations support both causal and noncausal types of superluminosity. For a Lagrangian of causal category, group velocity $v_g>d\omega/dk>c$ may display superluminal (SL) propagation.

In causal case, action of the source remains in the lightcone \cite{SusskindSuperluminal}, even though pulses exhibit superluminal propagation. The consequences of causal SL propagation are best demonstrated in the groundbreaking experiment \cite{NatureSLexp2000}. In this experiment, signal reaches the detector faster than the propagation of light in vacuum (c), after traversing a dye solution. However, a recent experiment \cite{TalukderPRL2014} showed that observed \cite{NatureSLexp2000} SL propagation is misleading in the sense that observed SL pulse peak is an analytical continuation of the eariler portion of the input pulse. In addition, we showed that \cite{metasginSL} velocity measurements calculated in Ref.~\cite{NatureSLexp2000} are not reliable on their correspondence to a real physical flow, within purely mathematical considerations. 

Different than the causal one, in the noncausal superluminosity case, action of a source can propagate outside the light cone \cite{SusskindSuperluminal}. In this case, solution at later times is not always determined by the initial (earlier) time. Superluminal communication of the source and the potentials \cite{PSpotentials} manifests itself with a refractive index $n(\omega)$ whose nonanalyticities shift to the upper-half of the complex-$\omega$ plane, see Sec.~\ref{sec:SLcommunication_causality}. Electric and magnetic fields exhibit noncausal connection for this type of refractive index. One knows that classical theory of electromagnetic fields complies with the special theory of relativity \cite{GriffithsEMT,JacksonEMT}. In parallel with the electromagnetic theory, order of events may change in superluminal reference frames according to special theory of relativity \cite{Tolman}.

Recent studies \cite{ScienceNews,entanWorm1,entanWorm2,entanWorm3} revealed an exciting connection between two theories working in two different regimes, general theory of relativity and quantum mechanics. Two quantum-entangled particles in 3-dimensions and two particles connected with a wormhole in 4-dimensions re equivalent descriptions of the same physics \cite{entanWorm1,entanWorm2,entanWorm3}. Therefore, action of one of the two entangled particles can propagate out of the light cone (even though uncontrolled \cite{nosuperluminal_communication}) in both pictures.

Discussions of Liberati and Maccione \cite{NatureNews,MaccionePRL2014} shifts the scientific curiosity to another medium, to the constituents of space-time. Observation of the absence of damping of high-energetic photons from Crab Nebula \cite{CrabNebula} signals the superfluid behavior \cite{MaccionePRL2014} of constituents (ensemble) of space-time, if they existed. In our recent study \cite{metasginHPtransformation}, we demonstrate that; nonclassicality of a single-mode photon field can be visualized as the entanglement of the vacuum (the constituents) generating these photons (as quasiparticle excitations). More explicitly, Holstein-Primakoff transformation \cite{Emary&BrandesPRE2003,Holstein&Primakoff} on the symmetric Dicke states   of an ensemble of identical particles \cite{PSidentical,symmetricDicke} maps the $N$-particle (collective) entangled states onto the single-mode nonclassical (e.g. squeezed) states. In this manner, if constituents of vacuum exists, they should be in an exchange symmetric state. We know from Bose-Einstein condensates (which are forced to be in symmetric states) that, interaction in such ensembles produces coherence (superfluidity), not damping (see p. 2 in Ref.~\cite{Pethick&Smith}). In Holstein-Primakoff mapping these interactions (entanglement) produce nonclassicality in the photon field.

In this paper, we demonstrate a further interesting connection. An optomechanical cavity produces nonclassical single-mode output beyond a critical cavity-mechanical coupling $g \gtrsim 2\sqrt{\gamma_m/\gamma_c}\: \omega_m$, with $\omega_m$ and $\gamma_m$ are resonance and damping rate of mechanical oscillator and $\gamma_c$ is the damping rate of the cavity. We raise and investigate the following question. One can place this optomechanical cavity as an optical component into a photonic device. What is the index $n(\omega)$ of the dielectric material replacing this optomechanical cavity in \textit{any} optical setup? So, we model the index of the optomechanical, for any frequency $\omega$, by examining the reflected/transmitted components of the incident light \cite{Agarwal2010,TarhanPRA2013,Boydbook}, see Fig.~\ref{fig1}. 

We reach a very interesting consequence. The nonanalyticity of $n(\omega)$ moves to the upper-half of the complex-$\omega$ plane at the same critical cavity-mechanical coupling, $g_{\rm crt} \simeq 2\sqrt{\gamma_m/\gamma_c}\: \omega_m$, where nonclassicality (e.g. squeezing) of the output light introduces, see Fig.~\ref{fig2}. Moreover, nature of existence of noncausal $n(\omega)$ is independent from the choice of length (boundary conditions \cite{PSboundary,MostafazadehPRL2013,MostafazadehPRL2009,SoukoulisPRB2002,SoukoulisPRB2005}) or type of type (two-sided or single-sided) of the cavity. We also observe that nonclassicality and noncausal behavior emerge at the same critical coupling, when $\gamma_m$ and $\gamma_c$ are varied.

We are aware that, in typical experiments on optomechanics \cite{optomechanicsEITexp} effective coupling $g$ is increased by deriving the field with a coupler laser. That is cavity is an active medium. On the other hand, in general, no physical rule restricts obtaining strengths of $g\gtrsim 2\sqrt{\gamma_m/\gamma_c}\:\omega_m$ without use of a coupler laser. Additionally, $g\gtrsim 2\sqrt{\gamma_m/\gamma_c}\:\omega_m$ (for effective detuning $\Delta=\omega_m$) corresponds to a perfectly stable regime of the optomechanical system, which is 3 orders small compared to instability $g>g_{\rm PT}\simeq \sqrt{\omega_m\Delta/2}$ \cite{optomechEnt,PSoptoPT}.

Occurrence of i) such a relation between quantum optical treatment and classical electromagnetic theory [$n(\omega)$], combined with ii) single-mode nonclassicality implies the entanglement of constituent identical (symmetric) particles at different positions \cite{metasginHPtransformation}, iii) entanglement is equivalent to a shortcut wormhole \cite{ScienceNews,entanWorm1,entanWorm2,entanWorm3}, iv) ideas of Ref.~\cite{NatureNews,MaccionePRL2014} on the structure of space-time, v) interaction of symmetrized particles do not lead to damping \cite{Pethick&Smith}, and vi) equivalence of noncausal behavior to the lift of barrier on faster-than-light communication of source and potentials (see Sec.~\ref{sec:SLcommunication_causality}); naturally leads one to raise the following question. Is nonclassicality the superluminal communication of different positions of vacuum due to entanglement? This behavior may lead to violation of causality among the electromagnetic fields on the reflecting surface of the cavity.

The paper is organized as follows. In Sec.~\ref{sec:reflection_transmission}, we introduce the basics of an optomechanical system. We obtain reflected and transmitted wave amplitudes via classical (c-number) version of Langevin equations, using the method introduced by Agarwal and Huang \cite{Agarwal2010,TarhanPRA2013,Boydbook}. In Sec.~\ref{sec:index} we determine the refractive index $n(\omega)$ describing the behavior of optomechanical cavity if it is used as an optical component. We show that emergence of noncausal behavior of the index is a phenomenon independent of the choice of length and the type of the cavity. In Sec.\ref{sec:SLcommunication_causality}, we briefly show that emergence of noncausal behavior, in classical electromagnetic theory, is followed by the lifting off the barrier on faster-than-light source-potential communication. In Sec.~\ref{sec:nonclassicality}, we investigate the quantum optical features of the output fields. In Sec.~\ref{sec:nonclass_measure}, we introduce the measure for the single-mode nonclassicality \cite{metasginNCmeasure}. In Sec.~\ref{sec:time_evolution}, we use the second-quantized Langevin equations and determine the time evolution of output fields, within the presence of noise. Hence, we determine the nonclassicality of output fields. In Sec.~\ref{sec:mutual_emergence}, we show that nonclassicality of the output fields and noncausal behavior of the linear response (refractive index) emerge at the same critical cavity-mechanical coupling for different values of $\gamma_m$ and $\gamma_c$. In Sec.~\ref{sec:conclusions}, we discuss the connections with other recent studies.

\section{Reflection and Transmission through an optomechanical cavity} \label{sec:reflection_transmission}
 
In this section, we briefly describe the physics and parameters of an optomechanical system. We introduce the second-quantized Hamiltonian governing the dynamics of the system. We obtain the quantum Langevin equations by introducing the input-output formalism (not noises) into equations of motion. In the present section, we are interested in the \textit{classical} and linear behavior of reflected and transmitted waves though the cavity. Hence, we replace the second-quantized operators with amplitudes (c-number). Second-quantized quantum approach will be carried in Sec.~\ref{sec:nonclassicality}, where we are interested in the nonclassical (entanglement) features of the fields.

For the purposes of our solutions to be more accessible (physically explicit), we use the input-output formalism in proper physical units, e.g. field operators are not in units of 1/$[{\rm frequency}]^{1/2}$ or coupling is not $\sqrt{\gamma_c}$, given in Chapter 9 of Ref.~\cite{Scullybook} and derived more explicitly in Ref.~\cite{metasginWEB}.

An optomechanical system consists of an optical cavity and a mechanical oscillator, placed into the cavity, with typical resonance frequencies ($\omega_m$) above MHz. The cavity field (frequency $\omega_c$) --driven by an external strong source ($\omega_{\rm L}$), also called coupler-- interacts with the mechanical oscillator via radiation pressure Hamiltonian $\hbar g_0 \hat{c}^\dagger\hat{c}\hat{x}_m$. Cavity-mechanical coupling introduces squeezing (nonclassicality) \cite{optomechEnt,optosqueezing_expr,optosqueezing_meystre} in the cavity and the output fields. Mirror(s) of the cavity are partially transparent with typical damping rates of $\gamma_c\approx 0.1\times \omega_m$. The damping of the mechanical oscillator $\gamma_m$ is in the order of Hz only, that is $\gamma_m\approx 10^{-6}\omega_m$.

In addition to the strong coupler field $\omega_L$, the cavity field is driven with a weak probe field of frequency $\omega_p$. The presence of the coupler field only for the purpose of increasing the effective coupling between the probe and the mechanical oscillator. The second-quantized Hamiltonian in the rotating frame with the coupler field frequency $\omega_{\rm L}$ can be written as \cite{Agarwal2010}
\begin{equation}
\hat{\cal H}=\hbar\Delta_c \hat{c}^\dagger\hat{c} + \hbar\omega_m\hat{a}_m^\dagger\hat{a}_m -\hbar g_0 \hat{c}^\dagger\hat{c} \hat{q}_m
+i\hbar g_c \alpha_{\rm L} (\hat{c}^\dagger-\hat{c}) \; ,
\end{equation}
where $\hat{c}$ and $\hat{a}_m$ are the annihilation operators for the cavity field and the phonon field of the mechanical oscillator, respectively. $\hat{q}_m=(\hat{a}_m^\dagger+\hat{a}_m)/\sqrt{2}$ is the displacement operator for the mechanical oscillator. $\Delta_c=\omega_c-\omega_{\rm L}$ and $\Delta_p=\omega_p-\omega_{\rm L}$ are the frequencies of cavity and probe fields in the rotating frame. $g_0=\omega_c\: q_0/L$ is the strength of cavity-mechanical coupling, with $a_0=\sqrt{\frac{\hbar}{m\omega_m}}$ is the harmonic oscillator length  for the mechanical oscillator. $|\alpha_{\rm L}|^2$ is the the number of photons in the strong coupler field and $g_c$ is the coupling of the cavity (via semitransparent mirror) to the coupler driving field. $g_c$ is related with the cavity damping as $\gamma_c=\pi D(\omega_c) g_c^2$, see Eq. (9.1.14) in Ref.~\cite{Scullybook} and Ref.~\cite{metasginWEB}, where $D(\omega_c)$ is the density of states at the cavity resonance $\omega_c$. Neither $g_c$ nor $D(\omega_c)$ will enter the final results of the calculations.

\begin{figure}
\includegraphics[width=3.5in]{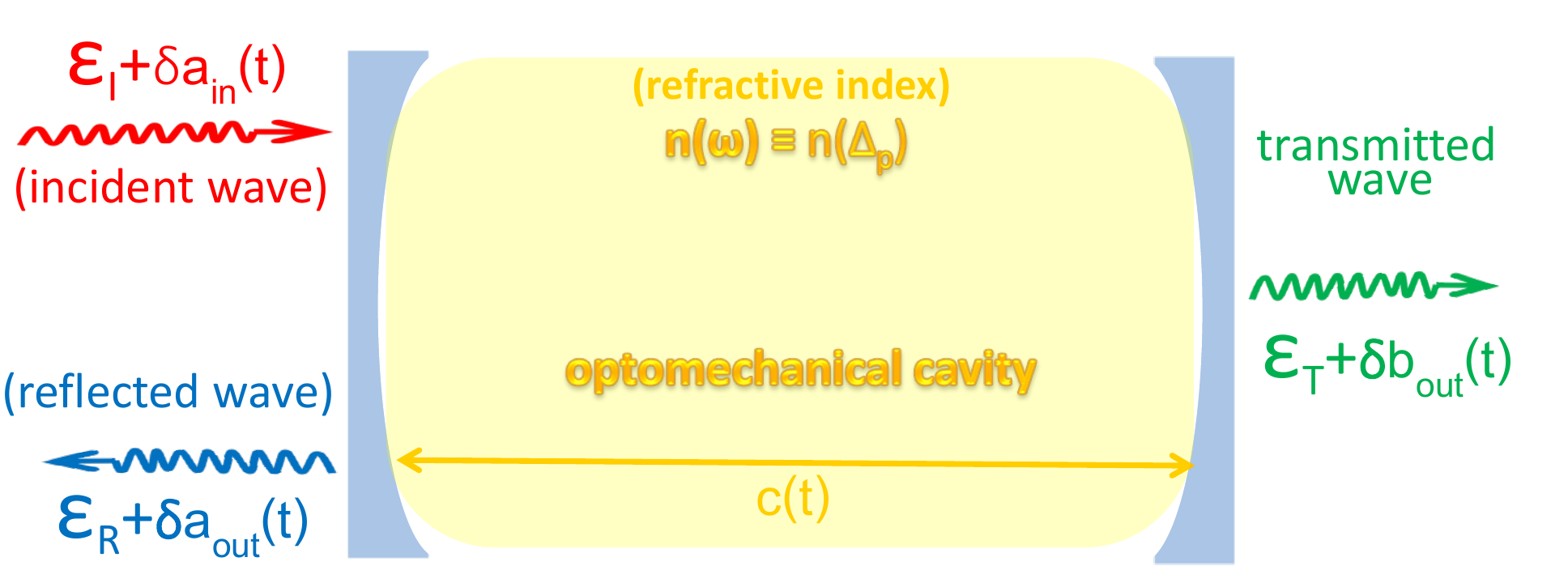}
\caption{A mechanical oscillator of frequency $\omega_m\approx$MHz is placed in an optomechanical cavity. Cavity has two semitransparent mirrors with damping rate $\gamma_c \approx 0.1\omega_m$ for each. Cavity is pumped with a strong coupler laser (not shown) to increase the effective cavity-mechanical coupling. The weak probe field, incident from the left side ($\varepsilon_I$), creates a reflected ($\varepsilon_R=\langle\hat{a}_{\rm out}(t)\rangle$) and a transmitted ($\varepsilon_T=\langle\hat{b}_{\rm out}(t)\rangle$) wave through the cavity. (a) In order to obtain information on the linear response of the cavity, we assign a refractive index $n(\omega)$ to the cavity. We find $n(\omega)$ using the reflected/transmitted waves determined from Langevin equations. Noncausal linear response emerges also for single-sided cavity, similarly at $g_{\rm crt}\approx 2\sqrt{\gamma_m/\kappa}$, due to the noncausal matching of the incident/reflected waves in the same medium (vacuum). (b) For investigating the quantum entanglement (nonclassicality) features of the reflected/transmitted fields, we include noises, $\delta\hat{a}_{in}$, $\delta\hat{a}_{out}$ and $\delta\hat{b}_{out}$, into the Langevin equations. Output fields become nonclassical at the same $g_{\rm crt}\approx 2\sqrt{\gamma_m/\kappa}$.}
\label{fig1}
\end{figure}

Including also the probe field, with $|\alpha_p|^2$ number of photons, Langevin equations take the form \cite{Agarwal2010,TarhanPRA2013}
\begin{subequations}
\begin{align}
&\dot{\hat{q}}_m = \omega_m \hat{p}_m \: , \label{Langevin_class_a} \\
&\dot{\hat{p}}_m=-\gamma_m \hat{p}_m -\omega_m \hat{q}_m + g_0 \hat{c}^\dagger\hat{c} \: , \\
&\dot{\hat{c}}=-(\kappa+i\Delta_c)\hat{c} + ig_0 \hat{q}_m \hat{c} + g_c\alpha_{\rm L} + g_c\alpha_p e^{-i\Delta_pt} \: , \label{Langevin_class_c}
  \end{align}
\end{subequations}
where probe field $\alpha_p$ is very small. Total damping rate is $\kappa=2\gamma_c$ for two-sided  and $\kappa=\gamma_c$ for single-sided semitransparent cavities, respectively. Reflected and transmitted waves (see Fig.~\ref{fig1}), induced by the probe field, can be determined by the input and output relations
\begin{subequations}
\begin{align}
\hat{a}_{\rm out}(t)=-\hat{a}_{\rm in}(t)+ 2\pi D(\omega_c)g_c \hat{c}(t) \: , \label{input_outout_a}\\
\hat{b}_{\rm out}(t)=2\pi D(\omega_c)g_c \hat{c}(t) \label{input_outout_b}\: ,
\end{align}
\end{subequations}
where $\hat{a}_{\rm out}(t) \equiv$ {\it reflected wave} ($\varepsilon_R$) and $\hat{b}_{\rm out}(t) \equiv$ {\it transmitted wave} ($\varepsilon_T$).

In determining the reflected/transmitted amplitudes for the incident probe field $\alpha_pe^{-i\Delta_pt}$, $Delta_p=\omega_p-\omega_{\rm L}$, we only consider the classical features of the fields. We replace $\hat{c} \to \alpha_c$, $\hat{a}_m \to \alpha_m$ and ignore the quantum aspects of the correlations, e.g. $\langle\hat{q}_m\hat{c}\rangle=\langle\hat{g}_m\rangle\langle\hat{c}\rangle$ \cite{Agarwal2010,TarhanPRA2013}. If one examines the steady states of the oscillations in Eq.s~(\ref{Langevin_class_a})-(\ref{Langevin_class_c}), he/she can find that $e^{-i\Delta_pt}$ and $e^{i\Delta_pt}$ oscilations are induced in the first order \cite{Agarwal2010,TarhanPRA2013,Boydbook} (for small $\alpha_p$) as
\begin{subequations}
\begin{align}
&\dot{q}_m = q_0 + q_+\alpha_p e^{-i\Delta_pt} + q_-\alpha_p^* e^{i\Delta_pt} \label{classical_sol_a} \: , \\
&\dot{p}_m = p_0 + p_+\alpha_p e^{-i\Delta_pt} + p_-\alpha_p^* e^{i\Delta_pt} \label{classical_sol_b} \: ,\\
&\dot{c} = c_0 + c_+\alpha_p e^{-i\Delta_pt} + qc-\alpha_p^* e^{i\Delta_pt} \label{classical_sol_c} \: , 
\end{align}
\end{subequations}
where $q_m(t)=[\alpha_m(t)^*+\alpha_m(t)]/\sqrt{2}$ and $q_m(t)=i[\alpha_m(t)^*-\alpha_m(t)]/\sqrt{2}$ are displacement and momentum for the mechanical oscillator. We note that $c_{0,\pm}$, $q_{0,\pm}$ and $p_{0,\pm}$ are all in dimensionless (scaled) form in Eq.s~(\ref{classical_sol_a})-(\ref{classical_sol_c}).

In order to obtain the linear response of the cavity, we insert Eq.s~(\ref{classical_sol_a})-(\ref{classical_sol_c}) into Eq.s~(\ref{Langevin_class_a})-(\ref{Langevin_class_c}), with $\hat{c}\to \alpha_c(t)$. We obtain equations for $c_{0,\pm}$, $q_{0,\pm}$ and $p_{0,\pm}$ by equating the terms oscillating with $e^{\pm i \Delta_p t}$ and nonoscillating terms. We ignore $e^{\pm i 2\Delta_p t}$, similar to Ref.s~\cite{Agarwal2010,TarhanPRA2013,Boydbook}, since we are interested in the linear response and $\alpha_p$ is very small. Reflected and transmitted waves, oscillating with probe frequency as $e^{- i \Delta_p t}$, are determined to be 
\begin{subequations}
\begin{align}
&\varepsilon_R(t)=[ -1 + 2\pi D(\omega_c)g_c c_+ ] \varepsilon_I \label{epsR} \\
&\varepsilon_T(t)= 2\pi D(\omega_c)g_c c_+ ] \varepsilon_I \label{epsT}
\end{align}
\end{subequations}
using the input-output relations in Eq.s (\ref{input_outout_a}) and (\ref{input_outout_b}), with dimensionless coefficient is 
\begin{equation}
c_+=\frac{\left( [\kappa-i(\Delta+\Delta_p)](\Delta_p^2-\omega_m^2+i\gamma_m\Delta_p) -i\omega_m |g|^2 \right) g_c }
{[(\kappa-i\Delta_p)^2+\Delta^2] (\Delta_p^2-\omega_m^2+i\gamma_m\Delta_p)+2\omega_m\Delta|g|^2}  \: .
\label{c+}
\end{equation}
Here, $g=g_0c_0$ is the effective coupling with $|c_0|^2$ is the number of photons in the cavity field at the steady state. Hence, presence of the strong coupler field ($\alpha_{\rm L}$) enhances the effective cavity-mechanical coupling several orders of magnitude in typical experiments \cite{optomechanicsEITexp}. Effective detuning $\Delta=\Delta_c-g_0q_0$ is obtained from steady-state equations. Damping rate is $\kappa=2\gamma_c$ and $\kappa=\gamma_c$ for double-sided and single-sided cavities, respectively. For the case of a single-sided cavity, $\varepsilon_T=0$ in Eq.~(\ref{epsT}) and $\varepsilon_R(t)$ remains unchanged in Eq.~(\ref{epsR}). 

It is important to note the following. In deriving Eq.~(\ref{c+}), using Eq.s~(\ref{classical_sol_a})-(\ref{classical_sol_c}), we ignored $e^{\pm i2\Delta_p t}$ terms. This corresponds to performing {\it rotating wave approximation} (RWA) in Eq.s~(\ref{Langevin_class_a})-(\ref{Langevin_class_c}).

\section{Corresponding refrective index $n(\omega)$} \label{sec:index}

In this section, we investigate the following question. One may place the optomechanical cavity in a photonic devices. We would like to know the linear response of this cavity. What is the refractive index $n(\omega)$ corresponding to this cavity, which generates the same reflected/transmitted pulses for a random probe field? We determine $n(\omega)$ using the continuity of electric and magnetic fields at the cavity interfaces. We reach an interesting result.  Nonanalyticity of $n(\omega)$ moves to the upper-half of the complex-$\omega$ plane when effective cavity-mechanical coupling exceeds $g=g_0 c_0 > 2\sqrt{\gamma_m/\kappa}\:\omega_m$. This implies the noncausal behavior of the fields.

Linear response of the cavity can be determined using four matching conditions for the fields. At the interfaces, parallel components of the total electric field $E_\parallel$ and the magnetic field $H_\parallel$ must bu continuous \cite{GriffithsEMT,JacksonEMT}, see Fig.~\ref{fig1}. Ratio of dielectric and magnetic permeabilities is determined to be 
\begin{equation}
\frac{\epsilon}{\mu} = \frac{-(\bar{\varepsilon}_R-1)^2+\tilde{\varepsilon}_T^2}{(\bar{\varepsilon}_R+1)^2-\tilde{\varepsilon}_T^2}
\label{epsmuratio1}
\end{equation}
where $\bar{\varepsilon}=\varepsilon_R /\varepsilon_I$ and $\tilde{\varepsilon}_T=\varepsilon_R e^{ikL} /\varepsilon_I$. Using Eq.~(\ref{epsR}), (\ref{epsT}) in Eq.~(\ref{epsmuratio1}) and introducing the dimensionless quantity $r=2\pi D(\omega_c)g_c$, numerator and denominator of $\epsilon/\mu$ becomes
\begin{equation}
\frac{\epsilon}{\mu} = \frac{-(rc_+-2)^2+ r^2c_+^2e^{i2kL}}{r^2(1-e^{i2kL})c_+^2} \: .
\label{epsmuratioc+}
\end{equation}

$\epsilon/\mu$ becomes noncausal if $c_+^2$, in the denominator, has a zero in the upper-half of the complex-$\omega$ plane. Since the denominator of $c_+$, given in Eq.~(\ref{c+}), is canceled in Eq.~(\ref{epsmuratioc+}), roots of the numerator of $c_+$
\begin{equation}
[\kappa-i(\Delta+\Delta_p)](\Delta_p^2-\omega_m^2+i\gamma_m\Delta_p)-i\omega_m|g|^2=0 \: ,
\label{rootsc+}
\end{equation}
that are $\Delta_p^{(1,2,3)}$, determine the positions of the nonanalyticities of $\epsilon/\mu$. If one of Im$\{\Delta_p^{(1,2,3)}\}$ is in the upper-half of the complex-$\omega$ plane, linear response of the cavity exhibits noncausal behavior. 

We note that emergence of noncausal behavior is independent from the length of the cavity ($L$) and wavevector of the incident probe, $k$. When the length of the model dielectric slab varies from the actual cavity size, noncausal behavior in $\epsilon/\mu$ [Eq.~(\ref{epsmuratioc+})] remains unchanged. Hence, noncausal behavior originates due to the temporal/frequency relations among the incident and reflected/transmitted waves, not due to boundary conditions \cite{MostafazadehPRL2013,MostafazadehPRL2009}. One can also show that same condition occurs similarly for index $n(\Delta_p)=\sqrt{\epsilon\mu}$, that is 
\begin{equation}
{\rm Im}\{\Delta_p^{(1,2,3)}\} > 0 \quad \Rightarrow \quad n(\Delta_p)\text{ is noncausal.}
\end{equation}

Interestingly, the same behavior is obtained by considering a single-sided cavity. In this cavity type, $\kappa=\gamma_c$, there is only incident and reflected waves through a semitransparent cavity mirror \cite{Aspelmeyer&Marquardt2014}. Similar to Eq.~(\ref{epsmuratioc+}), $c_+$ appears in the denominator of $\epsilon/\mu$. Therefore, noncausal behavior occurs at a single spatial position, that is on at the outer surface of the reflecting cavity.

In Fig.~\ref{fig2}a, we plot the imaginary part of the third root of Eq.~(\ref{rootsc+}), ${\rm Im}\{\Delta_{p}^{(3)}\}$, for different values of cavity-mechanical coupling $g=g_0c_0$. We observe that, for $g>g_c\approx 2\sqrt{\gamma_m/\kappa}$ nonanalyticity the refractive index moves to upper-half of the complex-$\omega$ plane. Hence, linear response of the cavity (output fields) becomes noncausal and the barrier on the faster-than-light source-potential communication is lifted off, see Sec.~\ref{sec:SLcommunication_causality}. Interestingly, Fig.~\ref{fig2}b reveals that at the same $g$, output field becomes single-mode nonclassical (see Sec.~\ref{sec:nonclassicality}). This implies the entanglement of the background particles --whose quasiparticle excitations the photons are-- which makes the instantaneous communication of different spatial positions possible.

\section{Superluminal source-potential communication and Violation of Causality} \label{sec:SLcommunication_causality}

In this section, we briefly summarize how a refractive index, with nonanalyticity in the upper-half of the complex-$\omega$ plane, allows faster-than-light communication of the source and potential.

If any of the response functions connecting the electric and magnetic fields, e.g. $D=\sqrt{\epsilon/\mu}B$ for Eq.~(\ref{epsmuratioc+}), has nonanalyticity in the upper-half plane; then the two fields has a noncausal connection \cite{JacksonEMT}. The amplitude of $D(t)$ is determined not only by the past, but feature of the $B(t')$ fields, $t'>t$. For example, $D(\omega)=\epsilon(\omega)E(\omega)$ leads to \cite{JacksonEMT}
\begin{equation}
D(t)=\int_{-\infty}^{\infty} dt' {\cal G}(t,t') E(t')
\label{D_E_relation}
\end{equation}
where response function 
\begin{equation}
{\cal G}(t,t')=\int_{-\infty}^{\infty} d\omega \epsilon(\omega) e^{-i\omega(t-t')}
\label{reponsefnx}
\end{equation}
must vanish for $t'>t$ in order to avoid violation of causality in Eq.~(\ref{D_E_relation}). For $t'<t$, Fourier term in Eq.~(\ref{reponsefnx}) becomes $e^{i\omega |\tau|}$, with $\tau=t-t'$ and $|\tau|$ positive. Integral in Eq.~(\ref{reponsefnx}) can be calculated by choosing the integration llop on the upper-half of the complex-$\omega$ plane \cite{JacksonEMT}, that is $\omega_I={\rm Im}\{\omega\}>0$, since $e^{-\omega_I|\tau|}$ vanishes on the infinite circle of the upper-half plane. Hence, if $\epsilon(\omega)$ has a nonanalyticity in the upper-half plane, violation of causality in the relation connecting the fields emerges.

Similarly, nonanalyticity of $\epsilon(\omega)$ or refractive index, implies the possibility of faster-than-light propagation of disturbnace from the source to potentials at distant points. Retarded Greens function \cite{JacksonEMT}
\begin{equation}
G^{(+)}({\bf r},t;{\bf r}',t') = \frac{1}{2\pi} \int_{-\infty}^{\infty} \frac{e^{ikR}}{R} e^{-i\omega \tau} d\omega
\label{Greensfnx}
\end{equation}
connects the source $\rho({\bf r}',t')$ and potential $\Phi({\bf r},t)$ as
\begin{equation}
\Phi({\bf r},t)=\int \int G^{(+)}({\bf r},t;{\bf r}',t') \rho({\bf r}',t') d^3{\bf r}' dt'
\end{equation}
with $R=|{\bf r}-{\bf r}'|$ and $\tau=t-t'$. Wavevector is related with refractive index as $k=n(\omega)\omega/c$. Analyticity of $n(\omega)$ in the upper-half of the complex-$\omega$ plane avoids the source-potential communications above the speed of the light, that is 
\begin{equation}
G^{(+)}({\bf r},t;{\bf r}',t') = G_0 \theta\left( \frac{\omega}{k}-\frac{|{\bf r}-{\bf r}'|}{|t-t'|}\right)
\end{equation}
with $\theta$ is the step function. However, if $n(\omega)$ has nonanalytical behavior in the upper-half of the complex-$\omega$ plane, disturbance of source can affect potential faster than the speed of light.

Therefore, faster-than-light communication of potential with the source and violation of causality among electric/magnetic fields, Eq.~(\ref{reponsefnx}) and Eq.~(\ref{Greensfnx}), have the same mathematical origin. This is the nonanalyticity of the refractive index (response functions) in the upper-half of the complex-$\omega$ plane. 

\section{Nonclassicality of the output fields} \label{sec:nonclassicality}

In the previous section, we examined the semi-classical behavior of reflected and transmitted waves through an optomechanical cavity. We arrived to the noncausal matching of the incident and reflected waves. In this section, we investigate the quantum optical features of reflected $\hat{a}_{\rm out}(t)$ and transmitted $\hat{b}_{\rm out}(t)$ waves. We determine the behavior of cavity $\hat{c}$ and output fields using the Langevin equations for noise operators. By deducing $\langle \hat{a}_{\rm out}^2(t)\rangle$ and $\langle \hat{a}_{\rm out}^\dagger (t) \hat{a}_{\rm out}(t)\rangle$ we determine the nonclassicality of the (e.g.) reflected field. The measure/criterion we use \cite{metasginNCmeasure} is adopted from Simon-Peres-Horodecki two-mode entanglement \cite{SimonPRL2000,AdessoPRA2004,WernerPRA2002} and both a necessary and sufficient criterion for Gaussian states \cite{optomechEnt}.

We fins the critical cavity-mechanical coupling where nonclassicality emerges. This comes out to be in the same value $g_{\rm crt}=2\sqrt{\gamma_m/\kappa}\: \omega_m$ with the emergence of noncausal behavior of reflected/transmitted waves. In other (and more explicit) words, single-mode nonclassicality emerges at the same critical coupling where the barrier avoiding the faster-than-light source-potential communication is lifted off.

\subsection{Single-mode nonclassicality measure} \label{sec:nonclass_measure}

Vogel and Sperling \cite{Vogel&Sperling2014,Mraz&Hage2014} showed that, i) the rank of two-mode entanglement, a single-mode field generates at the output of  a beam-splitter, is equal to the ii) number of terms needed to expand this single-mode state in terms of classical coherent states. In our recent study \cite{metasginNCmeasure}, we combine this result with the two-mode entanglement measure \cite{AdessoPRA2004,WernerPRA2002} for Simon-Peres-Horodecki \cite{SimonPRL2000} criterion. The very details and examples on squeezed states and superradiance can be found in Ref.~\cite{metasginNCmeasure}.

Since two-mode entanglement measure/criterion \cite{AdessoPRA2004,WernerPRA2002} is both a necessary and sufficient condition for Gaussian states, the single-mode measure/criterion is also a necessary and sufficient condition for Gaussain state. Optomechanical systems exhibit almost Gaussian states \cite{optomechEnt} when cavity-mechanical coupling strength is much below $g_{\rm PT}=\sqrt{\Delta \omega_m}$, where an instability takes place. Such transitions change the nature of the state abruptly \cite{metasginNCmeasure,Emary&BrandesPRE2003} from a Gaussian behavior.

We briefly summarize the measure \cite{metasginNCmeasure} for nonclassicality of a single-mode state. We use the beam-splitter transformations \cite{Kim&Knight2002,BSentanglement} to relate the elements of the two-mode covariance (noise) matrix $V_{ij}$, e.g. $V_{14}=\langle \hat{x}_1\hat{p}_2+\hat{p}_2\hat{x}_1\rangle/2$, to the noise elements of single-mode state, $\langle\hat{a}^2\rangle$ and $\langle\hat{a}^\dagger\hat{a}\rangle$. For example, off diagonal element is related as
\begin{equation}
V_{14}=\frac{1}{2}\langle \hat{x}_1\hat{p}_2+\hat{p}_2\hat{x}_1\rangle = -it^2\left(\langle\hat{a}^2\rangle e^{i2\phi} - \langle\hat{a}^2\rangle^* e^{-i2\phi}\right) \; .
\end{equation}
So as, we determine all elements of two-mode covariance matrix $v_{ij}$ in terms of only $\langle\hat{a}^2\rangle$ and $\langle\hat{a}^\dagger\hat{a}\rangle$. Once we determine the $V_{ij}$ matrix, we calculate the symplectic eigenvalues \cite{AdessoPRA2004,WernerPRA2002}
\begin{equation}
\eta^{\pm} \equiv \frac{1}{\sqrt{2}} \left( \sigma(V)  \pm \left\{  [\sigma(V)]^2-4{\rm det}(V)   \right\}^{1/2}   \right)^{1/2} \; ,
\label{eta}
\end{equation}
to find the logarithmic negativity 
\begin{equation}
E_{\cal N}={\rm max}(0,-\ln(2\eta^-)) \: .
\label{EN}
\end{equation}
$E_{\cal N}$ measures the degree of two-mode entanglement generated at the beam-splitter output by a single-mode input state,  $\langle\hat{a}^2\rangle$ and $\langle\hat{a}^\dagger\hat{a}\rangle$. The function $\sigma(V)$ in Eq.~(\ref{eta}) is calculated as $\sigma(V)=\det(V_m)+\det(V_c)-2\det(V_{mc})$, where $V_m$, $V_c$, $V_{mc}$ are 2$\times$2 matrices with $V_{ij}$ is written in the form
\begin{equation}
V = \begin{bmatrix}
V_m & V_{mc} \\
V_{mc}^T & V_c
\end{bmatrix} \: .
\label{V2by2form}
\end{equation}

\subsection{Time evolution of cavity output field} \label{sec:time_evolution}

We determine the time evolution of cavity noise operators ($\delta\hat{c}=\hat{c}-\langle\hat{c}\rangle$) using the standard methods \cite{optomechEnt,Tombesi1994} for the input-output formalism of an optomechanical cavity. Output field modes $\delta\hat{a}_{\rm out}(t)$ and $\delta\hat{b}_{\rm out}(t)$, see Fig.~\ref{fig1}, are related to the cavity mode $\hat{c}(t)$ as in Eq.s~(\ref{input_outout_a}), (\ref{input_outout_b}). Langevin equations for noise operators transform to \cite{optomechEnt}
\begin{subequations}
\begin{align}
&\delta\dot{\hat{q}}_m = \omega_m \delta\hat{p}_m \: , \label{Langevin_quantum_a} \\
&\delta\dot{\hat{p}}_m=-\gamma_m \delta\hat{p}_m -\omega_m \delta\hat{q}_m + g_0(\alpha_S^*\delta\hat{c} + \alpha_s\delta\hat{c}^\dagger) + 
g_m\hat{\epsilon}_{\rm in}(t) \label{Langevin_quantum_b} \\
&\delta\dot{\hat{c}}=-(\kappa+i\Delta_c)\delta\hat{c} + ig_0\alpha_s \delta\hat{q}_m + g_c\delta \hat{a}_{\rm in}(t)  
 \: , \label{Langevin_quantum_c}
\end{align}
\end{subequations}
where $\alpha_S=\langle\hat{c}\rangle$ is the cavity field amplitude at the steady-state and also equal to $\alpha_S=c_0$ which is introduced in Sec.~\ref{sec:reflection_transmission}. In fact, other steady-state values are also the same with Sec.~\ref{sec:reflection_transmission}, $q_s=\langle\hat{q}_m\rangle=q_0$ and $p_s=\langle\hat{p}_m\rangle=p_0$. Here, 
\begin{subequations}
\begin{align}
& \epsilon_{\rm in}(t)=\frac{i}{\sqrt{2}} \sum_{n} \left( \hat{a}_n^\dagger e^{i\omega_nt} - \hat{a}_n e^{-i\omega_nt} \right) \label{noisemech}\\
& \delta\hat{a}_{\rm in}(t)=-i\sum_{\bf k} e^{-i\omega_{\bf k}t} \hat{b}_{\bf k} \label{noiseopt}
\end{align}
\end{subequations}
are dimensionless \cite{Scullybook,metasginWEB} mechanical and vacuum noise operators whose expectations are
\begin{subequations}
\begin{align}
& \langle \epsilon_{\rm in}(t) \epsilon_{\rm in}(t')\rangle = \frac{\pi}{2} \rho(\omega_m)\delta(t-t') \label{noise_expectation_mech} \\
& \langle \delta\hat{a}_{\rm in}(t) \delta\hat{a}_{\rm in}(t')^\dagger \rangle = 2\pi D(\omega_c) \delta(t-t') \label{noise_expectation_opt}
\end{align}
\end{subequations}
with $\rho(\omega_m)$ and $D(\omega_c)$ are the density of states of phonons and photons, respectively, at the resonances. In Eq.~(\ref{Langevin_quantum_b}), $g_m$ is the coupling of the mechanical oscillator to the phonon reservoir and related to mechanical damping as $\gamma_m=\frac{\pi}{2}\rho(\omega_m)g_m^2$, similar to the cavity damping $\gamma_c=\pi D(\omega_c)g_c^2$.

Introducing the quadratures $\delta \hat{X}_c=(\delta\hat{c}^\dagger+\delta\hat{c})/\sqrt{2}$ and $\delta \hat{Y}_c=i(\delta\hat{c}^\dagger-\delta\hat{c})/\sqrt{2}$ and arrays $\hat{u}=[\delta\hat{q}_m \quad \delta\hat{p}_m \quad \delta\hat{X}_c \quad \delta\hat{X}_c ]^T$ and $\hat{u}_{\rm in}=[0 \quad g_m\epsilon_{\rm in} \quad g_c\delta\hat{X}_{\rm in} \quad g_c\delta\hat{X}_{\rm in} ]^T$, Langevin equations can be put into a matrix form
\begin{equation}
\hat{u}(t)=A\hat{u}(t) + \hat{u}_{\rm in}(t)
\label{uarray}
\end{equation}
with
\begin{equation}
A = \begin{bmatrix}
0 & \omega_m & 0 & 0 \\
-\omega_m & -\gamma_m & \sqrt{2}g_R & \sqrt{2}g_I \\
-\sqrt{2}g_I & 0 & -\gamma_c & \Delta \\
-\sqrt{2}g_R & 0 & -\Delta & \gamma_c  \\
\end{bmatrix} \: ,
\label{Amatrix}
\end{equation}
$g_R={\rm Re}\{g_0\alpha_s\}\equiv {\rm Re}\{g_0c_0\}$ and $g_I={\rm Im}\{g_0\alpha_s\}\equiv {\rm Im}\{g_0c_0\}$ and effective detuning is $\Delta=\Delta_c-g_0q_s$. Eq.~(\ref{uarray}) has the solution
\begin{equation}
\hat{u}(t)=M(t)\hat{u}(0) + \int_0^t dt' M(t') \hat{u}_{\rm in}(t-t')
\label{usol}
\end{equation}
where $M(t)=e^{At}$. If $|g|$ is in the stability regime, $|g| \lesssim \sqrt{\omega_m\Delta/2}\approx \omega_m/\sqrt{2}$ for $\Delta=\omega_m$ and $g\lesssim \sqrt{\kappa\gamma_m} \simeq 4\times 10^{-4}\omega_m$ for $\Delta=-\omega_m$ \cite{optomechEnt}, real parts of the eigenvalues of $A$ are all negative. Hence, for $t\to \infty$ the terms $e^{At}\hat{u}(0)$ in Eq.~(\ref{usol}) vanishes.

The noise of output field operators can be calculated, e.g. the reflected one,
\begin{equation}
\langle \hat{a}_{\rm out}^2(t) \rangle =\langle \left(\: -\delta\hat{a}_{\rm in}(t) 2\pi D(\omega_c) g_c \delta\hat{c}(t)\:\right)^2 \rangle \: .
\label{expextation_aout2}
\end{equation}
Hence, one needs the evaluation of second moments of the quadratures like
\begin{equation}
\langle \hat{X}(t)\hat{Y}(t)\rangle =\sum_{\ell_1=1}^4 \sum_{\ell_2=1}^4 \beta_{\ell_1 \ell_2} \int_0^t ds M_{3\ell_1}(t-s) M_{4\ell_2}(t-s)
\label{expectation_XY}
\end{equation}
are required as $t\to\infty$. In Eq.~(\ref{expectation_XY}), we defined $\langle \hat{u}_{{\rm in},\ell_1}(s)\hat{u}_{{\rm in},\ell_2}(s')\rangle=\beta_{\ell_1 \ell_2} \delta(s-s')$ where the Dirac-delta function of Eq.s~(\ref{noise_expectation_mech}), (\ref{noise_expectation_opt}) has dropped one of the integrals in Eq.~(\ref{expectation_XY}). Nonvanishing $\beta_{\ell_1\ell_2}$ values are, $\beta_{22}=\gamma_m$, $\beta_{33}=\beta_{44}=\gamma_c$ and $\beta_{34}=\beta_{43}^*=i\gamma_c$.

Fortunately, integral in Eq.~(\ref{expectation_XY}) can be calculated analytically using the linear algebra trick
\begin{equation}
A=PDP^{-1} \quad \Rightarrow \quad M(t)=e^{At}=Pe^{Dt}P^{-1} \:,
\end{equation}
where $P$ is the transformation diagonalizing ($D$) the evolution matrix $A$.

Placement of second moments --of the form of Eq.~(\ref{expectation_XY})-- into Eq.~(\ref{expextation_aout2}) shows that $\langle\delta\hat{a}_{\rm in}\delta\hat{c}\rangle = \langle\delta\hat{c} \delta\hat{a}_{\rm in}\rangle = 0$ vanishes for the {\it equal time} correlations. Hence, one obtains the relations 
\begin{equation}
\langle \hat{a}_{\rm out}^2(t) \rangle = 2\pi D(\omega_c) g_c \langle \delta\hat{c}^2(t)\rangle \: ,
\label{aout2_ct2}
\end{equation}
\begin{equation}
\langle \hat{a}_{\rm out}^\dagger(t) \hat{a}_{\rm out}(t) \rangle = 2\pi D(\omega_c) g_c \langle \delta\hat{c}^\dagger(t)\hat{c}(t)\rangle \: .
\label{naout_nc}
\end{equation}

Therefore, considering the necessary and sufficient single-mode nonclassicality criterion \cite{metasginNCmeasure} $|\langle\hat{a}^2\rangle|>\langle\hat{a}^\dagger\hat{a}\rangle$ --which is adopted from Duan-Giedke-Cirac-Zoller criterion \cite{Duan&Zoller2000}-- one can arrive that output fields and cavity field become nonclassical at the same critical coupling. In other words, the coefficient $2\pi D(\omega_c)g_c$, in Eq.s~(\ref{aout2_ct2}) and (\ref{naout_nc}), does not affect the nonclassicality feature.

To summarize: i) For determining the degree of nonclassicality we only need $\langle \hat{a}_{\rm out}^2(t) \rangle$ and $\langle \hat{a}_{\rm out}^\dagger(t) \hat{a}_{\rm out}(t) \rangle $. ii) $\hat{a}_{\rm out}$ depends on $\hat{c}(t)$, so we obtain the time evolution of the cavity field $\hat{c}(t)$ using standard form for (noise-added) Langevin equations \cite{optomechEnt,Tombesi1994}. iii) We relate second moments of $\hat{a}_{\rm out}$ to $\hat{c}(t)$ as given in Eq.s~(\ref{aout2_ct2}), (\ref{naout_nc}). iv) We calculate the second moments at $t \to \infty$ by Eq.~(\ref{usol}) \cite{optomechEnt}. v) We put these two moments into the covariance matrix (\ref{V2by2form}) and calculate the degree of single-mode nonclassicality using Eq.~(\ref{EN}). The inconvenience in using a parametric pump approximation is, the emergence and degree of the entanglement becomes sensitive to the phase of the coupler field $\alpha_{\rm L}=|\alpha_{\rm L}|e^{i\theta}$, see Eq.~(13) in Ref.~\cite{metasginEndfire} and Ref.s~\cite{parametricPhase1,parametricPhase2,parametricPhase3}. Hence, we also maximize $E_{\cal N}$ with respect to $\theta$, as described in figure 1 of Ref.~\cite{metasginNCmeasure}.

\section{Mutual emergence of nonclassicality and violation of causality} \label{sec:mutual_emergence}

In Sec.~\ref{sec:nonclassicality}, we presented the method for determining the nonclassicality degree of cavity output modes. In Sec.~\ref{sec:reflection_transmission}, we investigated the linear response of this optomechanical cavity by examining the incient and reflected waves at the front interface of the cavity. (Please note that, in a single-sided cavity noncausal behavior emerges with the same conditions.) In this section, we compare the behavior of nonclassicality degree $E_{\cal N}$ and noncausal behavior of the refractive index $n(\omega)$ for the same cavity-mechanical coupling $g=g_0c_0$.

\begin{figure}
\includegraphics[width=3.5in]{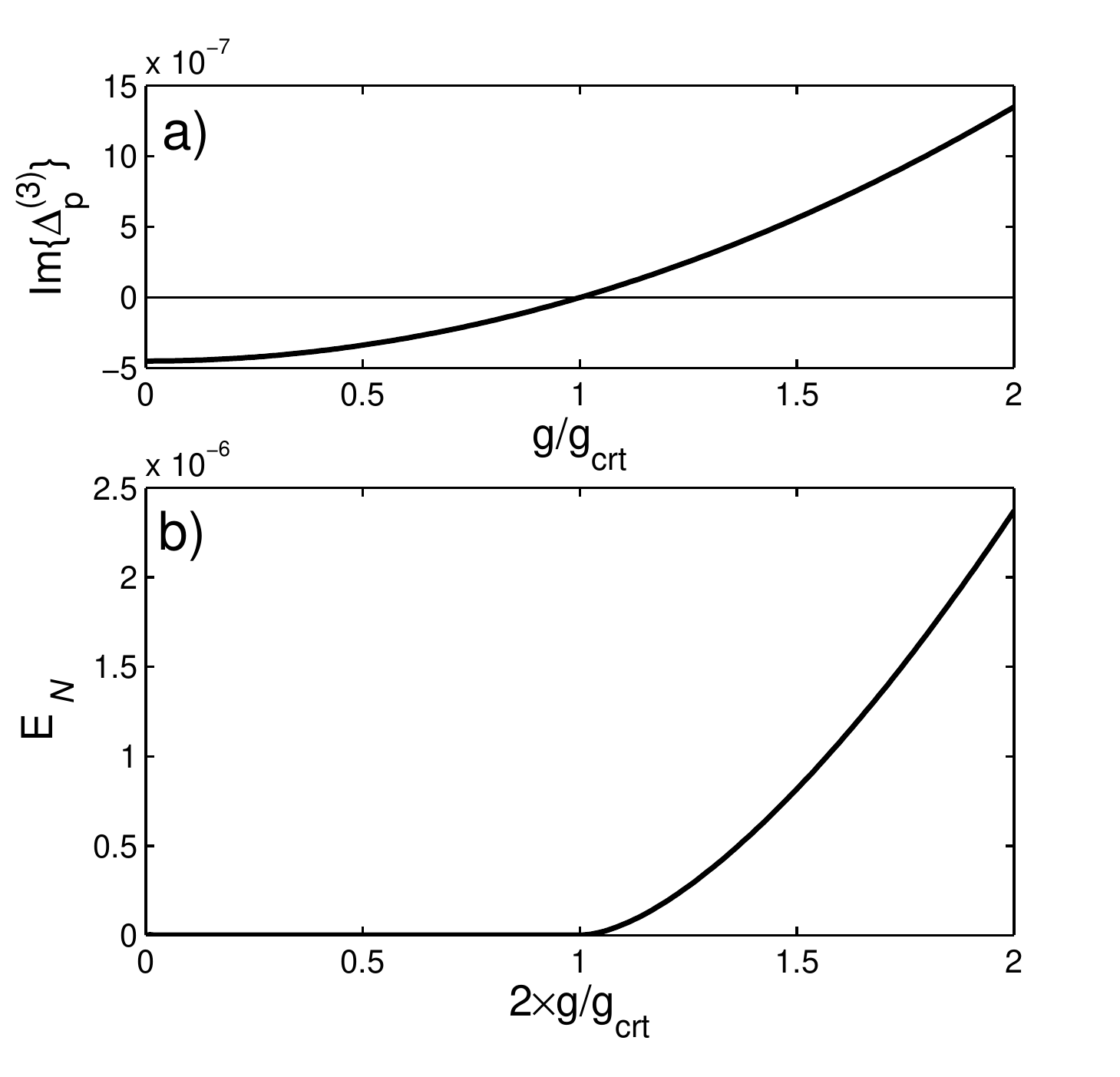}
\caption{(a) Noncausal behavior in the linear optical response of an optomechanical cavity. One of the nonanalyticities of the response function (index) moves to the upper-half of the complex-$\omega$ plane if cavity-mechanical coupling exceeds $g>g_{\rm crt}^{\rm (cls)}\approx 2\sqrt{\gamma_m/\kappa}\:\omega_m$, see Eq.~\ref{rootsc+}. Here, $\Delta_p$ is the probe frequency (incident wave) in the rotating frame. (b) The optomechanical output fields (also reflected/transmitted waves) become nonclassical at the same critical cavity-mechanical coupling. $g_{\rm crt}^{\rm (ncls)}=g_{\rm crt}^{\rm (ncls)}/2$, because a rotating wave approximation is performed in classical treatment, Sec.~\ref{sec:reflection_transmission}, with the neglectance of $e^{\pm i2\Delta_pt}$ terms. Therefore, the barrier on the faster-than-light communication, Sec.~\ref{sec:SLcommunication_causality}, is lifted at the same place where nonclassicality of the fields emerge. Single-mode nonclassicality can also be visualized as \cite{metasginHPtransformation} the entanglement of identical particles generating photons as quasiparticle excitations. $g_{\rm crt}^{\rm (ncls)}/2$ and $g_{\rm crt}^{\rm (cls)}$ come out to be equal for different values of $\gamma_m$ and $\kappa$. }
\label{fig2}
\end{figure}

In Fig.~\ref{fig2}, we plot the steady-state behavior of the nonclassicality degree $E_{\cal N}(t=\infty)$ for increasing values of cavity-mechanical coupling $g=g_0c_0$. We observe that, for $g>g_{\rm crt}^{\rm (cls)} \approx 2\sqrt{\gamma_m/\kappa}\:\omega_m$ (with $g_{\rm crt}^{\rm (cls)}$ determined empirically) output field becomes nonclassical. Surprisingly, at the same critical coupling, $g_{\rm crt}^{\rm (ncls)}=g_{\rm crt}^{\rm (cls)}$, the imaginary part of one of the 3 roots of $c_+=0$, Eq.~(\ref{rootsc+}), moves to the upper-half of the complex-$\omega$ ($\Delta_p$ here) plane, see Fig.~\ref{fig2}b. As can be seen from Eq.~(\ref{epsmuratioc+}), if $c_+$ has a nonanalyticity in the upper-half of the complex-$\omega$ plane, the response function $\epsilon/\mu$ (and also refractive index) exhibits noncausal behavior. This removes the barrier on the faster-than-light communication of the source and potential, see Sec.~\ref{sec:SLcommunication_causality}.

Hence, single-mode nonclassicality (collective entanglement of identical particles generating the single-mode quasi-excitations \cite{metasginHPtransformation}) and possibility of faster-than-light communication in classical electromagnetism emerges at the same same critical coupling. In Fig.~\ref{fig3}, we show that the two couplings, $g_{\rm crt}^{\rm (cls)}$ and $g_{\rm crt}^{\rm (ncls)}$ emerges at the same place for different $\gamma_m$ and $\gamma_c$.

In Fig.~\ref{fig2}b, we multiply the cavity-mechanical coupling by a factor of 2. Because, for the sake of deducing the linear response, see Sec.~\ref{sec:reflection_transmission} and Ref.s~\cite{Agarwal2010,TarhanPRA2013,Boydbook}, we omitted the oscillations $e^{-i2\Delta_p t}$ which occurs due to nonlinearities induced by cavity-mechanical interaction. This is equivalent to performing a \textit{rotating wave approximation} (RWA) on the small oscillations $c_\pm$, $q_\pm$ and $p_\pm$ \cite{PS_RWA}.

Is is a well known situation that a Hamiltonian on which a RWA is performed display the same critical phenomenon for 2 times stronger coupling compared to a Hamiltonian without RWA, see ${\rm 2}^{\rm nd}$ paragraph in the second page of Ref.~\cite{Emary&BrandesPRE2003} and Ref.s~\cite{RWA_SR1,RWA_SR2}. Comparison of classical and quantum (entanglement) features poses fundamental difficulties. In order to obtain the linear (classical) response in Sec.~\ref{sec:reflection_transmission} one needs to ignore $e^{-i2\Delta_p t}$ oscillations, or deal with them as nonlinear polarization. Hence, we cannot mimic the time evolution matrix $A$ in Eq.~\ref{Amatrix}. One can try to parform a RWA (for $\delta\hat{c}$, $\delta\hat{a}_m$) in the quantum picture, in Eq.s~(\ref{Langevin_quantum_a})-(\ref{Langevin_quantum_c}), in order to mimic the classical approach of Sec.~\ref{sec:reflection_transmission}. However, thi time quantum optical nature of the Hamiltonian changes, and nonclassicality of cavity/output modes never observed however much the strong the $g$ is within the stability regime \cite{PS_Hamlt_nature}. 

In Fig.~\ref{fig2}b, $\Delta=\omega_m$ and the system is in perfectly stable regime. For comparison, $g_{\rm crt}\approx 4\times 10^{-3}\omega_m$ and the critical coupling for the induction of instability is $g_{\rm PT}^{(+)}\approx \sqrt{\Delta \omega_m/2} \approx 0.7\omega_m$ \cite{optomechEnt}, with $\gamma_m=10^{-6}\omega_m$ and $\gamma_c=0.1\omega_m$ as in experiments \cite{optomechanicsEITexp}.

We also investigate the regime $\Delta=-\omega_m$, where instability of the solutions occurs at a very small cavity-mechanical coupling that is $g_{\rm PT}^{(-)}=\sqrt{\gamma_m\kappa}\approx 4\times 10^{-4}\omega_m$ \cite{optomechEnt}. Interestingly, the imaginary part of the root of $c_+=0$ (nonanalyticity of index) in Eq.~(\ref{rootsc+}), ${\rm Im}\{\Delta_p^{(3)}\}$, moves to the upper-half of the complex-$\omega$ plane at this critical value  $g_{\rm PT}^{(-)}$. However, we cannot perform simulations above $g> g_{\rm PT}^{(-)}$, for $\Delta=-\omega_m$ case, since both classical [Eq.s (\ref{classical_sol_a})-(\ref{classical_sol_c})] and quantum solutions [Eq.~\ref{usol}] become unstable in this regime. Nevertheless, we observe that output fields do not become nonclassical in the stable regime $g<g_{\rm PT}^{(-)}$.

\section{Conclusions} \label{sec:conclusions}

We show that emergence of nonclassical single-mode light from an optomechanical cavity is followed by noncausal behavior in the linear response of the cavity. The nature of emergence of noncausal behavior is independent from the length (boundary conditions) and the type of the cavity. Anomaly in the linear response emerges due to the noncausal matching of the incident and reflected waves at the front interface of the cavity. Emergence of noncausal behavior is shown to be strongly related with the possibility of faster-than-light communication between the source and potential.

The result of the present paper is intimately related with the outcomes of the following studies. Holstein-Primakoff mapping for nonclassical fields \cite{metasginHPtransformation}, wormhole representation of entanglement in 4D space \cite{ScienceNews,entanWorm1,entanWorm2,entanWorm3}, discussion on the constituents of the vacuum \cite{NatureNews,MaccionePRL2014}, mathematical equivalence of faster-than-light source-potential communication to violation of causality among fields (Sec.~\ref{sec:SLcommunication_causality}) and superluminal reference frames in Relativity \cite{Tolman}.

If vacuum indeed has constituents \cite{NatureNews,MaccionePRL2014}, ensemble of these constituents must be superfluid, since high-energetic photons from Crab Nebula reaches the observers undamped \cite{CrabNebula}. Holstein-Primakoff mapping of $N$-particle ($N$ large) states to single-mode photon states \cite{metasginHPtransformation} restricts the particles to occupy the (exchange) symmetric Dicke states \cite{PSidentical,symmetricDicke}. Hence, if vacuum has constituents --generating photons as quasiparticles-- these constituents are restricted to symmetrized states. We already know that collisions (interactions) in symmetrized particles (such as Bose-Einstein condensates) are responsible for the induction of superfluidit, rather than creating damping, see p.~2 in Ref.~\cite{Pethick&Smith}. 

The emergence of violation of causality --due to faster-than-light communication via entanglement-- may also be related with the theory of relativity. Maxwell equations are consistent  with relativity \cite{GriffithsEMT,JacksonEMT}. In relativity, for a reference frame moving with $v>c$, the order of events may change \cite{Tolman}, resulting in violation of causal behavior. Even though the fields (electric/magnetic and gravitational) are the quantities we observe in typical experiments, potentials are more fundamental quantities in quantum mechanics, see discussion in Chapter 2.6 of Ref.~\cite{Sakuraibook}.

%

\begin{acknowledgements}
M.E.T. thanks G\"ursoy B. Akg\"u\c{c} for illuminating and leading discussions, G\"on\"ul \"Unal and Se\c{c}kin K\"urk\c{c}\"uo\u{g}lu for courses on Field Theory, Bayram Tekin for discussions on General Theory of Relativity, Olcay Co\c{s}kun and Asl{\i}  Pekcan for mathematical support, Ceyhun Bulutay and Haluk Utku for motivational support. M.E.T. acknowledge support from T\"{U}B\.{I}TAK-KAR\.{I}YER  Grant No.  112T927 and T\"{U}B\.{I}TAK-1001  Grant No.  114F170. M.E.T. is supported by  Hacettepe University BAP-6091 Grant No. 014G602002.
\end{acknowledgements}


%
%
%

\begin{thebibliography}{99}
\bibitem{PSpotentials} One shall note that in quantum mechanics electromagnetic potentials are fundamental quantities --which may produce interferences-- even  though we observe the fields, see discussion in Chapter 2.6 in Ref.~\cite{Sakuraibook}.
\bibitem{PSidentical} Internal states of identical particles, both bosons and fermions, can accommodate only the symmetric set of Dicke states \cite{symmetricDicke}. This is because, there is no anti-symmetric set of Dicke states. External degree (motional states) of fermions are anti-symmetric.
\bibitem{PS_RWA} Please note that such a neglectance is not performed on nonoscillating terms $c_0$, $q_0$ and $p_0$. So, they are values without RWA $c_s$, $q_s$ and $p_s$. In other words, a RWA is carried on the effective Hamiltonian governing small fluctuations in Sec.~\ref{sec:reflection_transmission}, but not on the whole Hamiltonian.
\bibitem{PS_Hamlt_nature} It is quite interesting that for the interaction of two system with finite dimensions \cite{metasginSpinsqueezing}, the interaction Hamiltonian with RWA creates squeezing while interaction without RWA results no spin-squeezing.
\end{thebibliography}
\end{document}